\documentclass[a4paper,fleqn,usenatbib]{mnras}

\usepackage[T1]{fontenc}
\usepackage{ae,aecompl}

\usepackage{graphicx}	
\usepackage{amsmath}	
\usepackage{amssymb}	
\usepackage{subcaption}
\usepackage{cases}
\usepackage{multirow}
\usepackage{natbib,hyperref}

\title[Elliptical galaxies]{The effects of the IMF on the chemical evolution of elliptical galaxies.}

\author[C. De Masi et al.]{
Carlo De Masi,$^{1}$\thanks{E-mail: demasi@oats.inaf.it}
F. Matteucci,$^{1,2,3}$
F. Vincenzo,$^{4}$
\\
$^1$ Astronomy Department, University of Trieste, Via Tiepolo 11, I-34127 Trieste, Italy\\
$^2$ I.N.A.F. Osservatorio Astronomico di Trieste, via G.B. Tiepolo 11, I-34131, Trieste, Italy\\
$^3$ I.N.F.N. Sezione di Trieste, via Valerio 2, 34134 Trieste, Italy\\
$^4$ Centre for Astrophysics Research, School of Physics, Astronomy and Mathematics, University of Hertfordshire,\\ College Lane, Hatfield, AL10 9AB, UK\\
}

\date{Accepted XXX. Received YYY; in original form ZZZ}

\pubyear{2015}

\begin{document}
\label{firstpage}
\pagerange{\pageref{firstpage}--\pageref{lastpage}}
\maketitle

\begin{abstract}
We describe the use of our chemical evolution model to reproduce the abundance patterns observed in a catalog of elliptical galaxies from the SDSS DR4. The model assumes ellipticals form by fast gas accretion, and suffer a strong burst of star formation followed by a galactic wind which quenches star formation. Models with fixed IMF failed in simultaneously reproducing the observed trends with the galactic mass.\\
So, we tested a varying IMF; contrary to the diffused claim that the IMF should become bottom heavier in more massive galaxies, we find a better agreement with data by assuming an inverse trend, where the IMF goes from being bottom heavy in less massive galaxies to top heavy in more massive ones. This naturally produces a downsizing in star formation, favoring massive stars in largest galaxies. Finally, we tested the use of the Integrated Galactic IMF, obtained by averaging the canonical IMF over the mass distribution function of the clusters where star formation is assumed to take place. We combined two prescriptions, valid for different SFR regimes, to obtain the IGIMF values along the whole evolution of the galaxies in our models. Predicted abundance trends reproduce the observed slopes, but they have an offset relative to the data. We conclude that bottom-heavier IMFs do not reproduce the properties of the most massive ellipticals, at variance with previous suggestions.\\
On the other hand, an IMF varying with galactic mass from bottom-heavier to top-heavier should be preferred.
\end{abstract}




\section{Introduction}

The origin and evolution of elliptical galaxies has been long debated, in an attempt to explain the existence of a certain number of common properties.\\
To name a few, we remember the so-called fundamental plane (\textbf{FP}), i.e the tight correlation existing between the central velocity dispersion, the surface brightness and the effective radius of ellipticals \citep{dressler1987,djorgovski1987,bender1992,jorgensen1996,burstein1997}; the mass-metallicity relation (\textbf{MZR}), i.e. the trend of increasing absorption metal lines strength with galactic velocity dispersion, usually interpreted as an increase of the metal content in more massive galaxies \citep{lequeux1979,garnettshields1987,zaritsky1994h,garnett2002luminosity,pilyugin2004,tremonti2004origin,kewley2008metallicity,mannucci2010fundamental}; the color-magnitude relation (\textbf{CMR}), i.e. the observed reddening of higher mass galaxies \citep{bower1992}.\\
Two main scenarios have been proposed to model the formation of elliptical galaxies.\\
On one hand, models based on the hierarchical clustering of dark matter haloes picture ellipticals as the result of several merging events of spiral galaxies \citep{kauffmann1993,kauffmann1998}, with the consequence that more massive ellipticals should be formed at a lower redshift.\\
On the other hand, there are models based on the assumption that elliptical galaxies form following the fast, monolithic collapse of a gas cloud; the increased density resulting from such a collapse leads to a period of intense star formation, until the onset of a galactic wind, powered by the thermal energy injected into the ISM by SNe and stellar winds, drives the remaining gas away, thus quenching star formation.\\
\cite{larson1974} was one of the firsts to explore the application of the monolithic collapse scenario to elliptical galaxies, and he was able to reproduce the MZR and the CMR by assuming the star formation efficiency in the galaxy was inversely proportional to the total mass. Combined with the increasing depth of the gravitational potential well in more massive galaxies, this led to a later onset of the galactic wind (``classic" wind models), thus resulting in a prolonged period of star formation which naturally accounted for the presence of more metals, and the consequent reddening of the galaxy.\\
This theoretical framework had to be reconsidered following the observation of supersolar $[\alpha/Fe]$ ratios in the core regions of elliptical galaxies \citep{worthey1992, carollo1993, davies1993} and of the trend of increasing of this same ratio in more massive galaxies \citep{worthey1992, weiss1995, kuntschner2000}; these features hinted at a different scenario, where star formation lasts for a shorter period of time in more massive galaxies (``downsizing'' in star formation), thus preventing the stellar chemical composition to be polluted by Fe-peak element produced on a longer timescale by type Ia SNe.\\
To conciliate the $[\alpha/Fe]$ behavior with the MZR, \cite{matteucci1994} devised a modification of the classic wind model, where more massive galaxies form stars more efficiently (``inverse" wind models); due to this effect, massive galaxies undergo a period of star formation which is both intense enough to account for the increased metal content, and short enough to prevent the Fe-peak elements pollution of the stellar population.\\
Numerical models based on the inverse wind scenario have proved so far to be the best way to naturally account for both the MZR and the $[\alpha/Fe]$-mass ($\sigma_0$) relationships simultaneously in ellipticals \citep{P04}.\\
On the other hand, reproducing the effect of the downsizing formation on chemical abundances has been the main limit of cosmological formation models, where additional physical mechanisms are required to produce the initial burst and the following quenching of star formation.\\
\cite{pipino2008} attempted to reproduce the observed trends by implementing detailed treatments for the chemical evolution in semi-analytical models; taking into account the quenching effect on star formation driven by AGN feedback yielded $[\alpha/Fe]$ ratios marginally consistent with data, but they failed in reproducing the MZR.\\
\cite{calura2009,calura2011} presented a cosmological galaxy formation model including fly-by triggered starbursts combined with the late quenching effect of AGN feedback, which allowed them to better shape the produced $[\alpha/Fe]$-mass relationship. \cite{fontanot2017} studied the application of the IGIMF in the context of the semy-analytical model GAEA, with the result of reproducing the observed increase of $\alpha$-enhancement with stellar mass in the considered datasets, whereas using a canonical, universal IMF yielded a flatter slope of the relation.\\
The aim of this paper is to reproduce the observed abundance patterns in a large sample of elliptical galaxies, by adopting a revised version of the Pipino \& Matteucci (2004) model assuming different prescriptions for the IMF.
This work is organized as follows.\\
In Section \ref{sec:catalog_T10}, we describe the dataset used for the comparison with our chemical evolution models.\\
In Section \ref{sec:chem_evol_models}, the chemical evolution model we used is presented; we detail the basic equations and principles of the model, introduce the treatment of the energetics and describe how to relate the model predictions to the observable quantities in the dataset.\\
Finally, Section \ref{sec:results} is dedicated to the results. We first examine the different outputs provided by 
simply varying only the basic parameters of the models, always within the downsizing formation scenario; then, we extend the analysis, by including variations of the IMF in the physical picture. Finally, we test and discuss the effects of taking into account the Integrated Initial Mass Function (IGIMF).

\section{Catalog}\label{sec:catalog_T10}
The dataset used to test our models is the one originally presented in \cite{thomas2010}.\\
It consists of 3360 early-type galaxies in the redshift range 0.05 < z < 0.06, extracted from the MOSES (Morphologically Selected Early types in SDSS) catalogue, which have been morphologically inspected and classified as early-types out of an initial sample extracted from the Data Release 4 (DR4) of the Sloan Digital Sky Survey (SDSS).\\
The use of a morphological selection criteria had the aim of removing any bias against recent star formation; as a result, the dataset has been further divided into two separate subsets:
\begin{itemize}
	\item a subset of objects with ages peaked at old values, analogous to the ``red sequence'' observed in the color-magnitude of populations including both late and early-type galaxies \citep{devaucouleurs1961,strateva2001,bell2004};
	\item a small $(\approx 10\%)$ fraction of galaxies with ages peaking at $\approx 2.5\,Gyr$, analogous to the so-called ``blue cloud''. The lower ages, lower $[\alpha/Fe]$ ratio values and the presence of signs of star formation suggest these galaxies have been rejuvenated by means of recent, minor star formation events. 
\end{itemize}
For the comparisons with our models, we always only used the galaxies in the ``red sequence'' subset of the catalog.\\
Objects in the catalog had their stellar population parameters (luminosity-weighted ages, total metallicities, $[\alpha/Fe]$ ratios) derived from the fitting of the 25 Lick absorption line indices to spectro-photometric models \citep{thomas2003stellar}; moreover, an estimate of the dynamical mass is provided, which should also provide a good approximation of the galactic baryonic mass inside the effective radius \citep{cappellari2006,thomas2007}.\\
As for the $[\alpha/Fe]$ ratio, in \cite{thomas2003stellar} the following relationship between total metallicity [Z/H], $[\alpha/Fe]$ and iron abundance [Fe/H] is assumed to hold:
\begin{equation}\label{eq:AFE_definition}
[Z/H] = [Fe/H] + A[\alpha/Fe]
\end{equation}
with $A=0.94$.\\
To account for possible differences in the $\alpha$-elements assumed to contribute to the total $[\alpha/Fe]$ ratio in the models and in the data, throughout the text we compared the $[\alpha/Fe]$ ratio provided in the dataset both to our predicted $[Mg/Fe]$ and to an $[\alpha/Fe]$ estimate derived from our model by inverting equation \ref{eq:AFE_definition}.

\section{Chemical evolution model}\label{sec:chem_evol_models}

The chemical evolution model for elliptical galaxies used is the one presented in \cite{P04} (hereafter, P04), and it allows for:
\begin{itemize}
	\item multi-zone representation of the galaxy, described as formed by spherical non-interacting shells with fixed thickness of $0.1R_{eff}$;
	\item taking into account the presence of Type I/II SN feedback;
	\item taking into account the effect of an initial infall episode;
\end{itemize}
For each shell, the evolution of the i-th element abundance is described by solving the equation of chemical evolution \citep[CEQ -][]{matteucci1986,matteucci1995}:
\begin{equation}\label{CEQ}
\begin{split}
\frac{dG_{i}(t)}{dt} &= -\psi(t)\,X_{i}(t) + \\ +\int_{M_L}^{M_{Bm}}\psi(t-\tau_{m})\,&Q_{mi}(t-\tau_m)\,\varphi(m)\,dm\, +\\
+ A\,\int_{M_{B_m}}^{M_{B_M}}dm\,\varphi(m)&\left[ \int_{\mu_{m}}^{0.5}f(\mu)\,\psi(t-\tau_{m_2})\,Q_{mi}(t-\tau_{m_2})d\mu \right]+\\
+(1-A)\int_{M_{B_m}}^{B_M}&\,\psi(t-\tau_{m})\,Q_{mi}(t-\tau_m)\,\varphi(m)\,dm+\\
+\int_{M_{B_M}}^{M_U}\,\psi(t-\tau_{m})\,&Q_{mi}(t-\tau_m)\,\varphi(m)\,dm+\\
+\left[ \frac{dG_i(t)}{dt} \right]&_{infall}
\end{split}
\end{equation}
where the four integrals represent the restitution rate of the i-th chemical element to the ISM due to stars in different mass ranges: single stars with mass between $M_L=0.8\,M_{\odot}$ and $M_{B_m}=3\,M_{\odot}$; binary systems, ending their lives as type Ia SNe, with total mass of the system comprised between $M_{B_m}=3\,M_{\odot}$ and $M_{B_M}=16\,M_{\odot}$; single stars in the same mass range, i.e. stars dying either as C-O dwarfs (up to $8\,M_{\odot}$) or type II SNe; stars with mass larger than $16\,M_{\odot}$, producing core collapse (either type II or Ib/c) SNe.\\
In the equation:
\begin{itemize}
	\item the abundance by mass of the i-th chemical species in the ISM $X_i(t)$ is defined as:
		\begin{equation*}
		X_i(t) \equiv \frac{M_{i}}{M_{gas}}
		\end{equation*}
	with the condition:\\
		\begin{equation*}
		\sum_{i=1}^{N}X_{i}=1
		\end{equation*}
	\item $G_{i}(t)$ is the ratio between the mass density of the element i at the time t and its initial value
		\begin{equation}
		G_{i}(t) = X_{i}(t)\left( \frac{\rho_{gas}(t)}{\rho_{gas}(0)}\right) 
		\end{equation}
	\item the star formation rate $\psi(t)$ is assumed to follow the law
		\begin{equation}
		\psi(t)=	
		\begin{cases}
		\displaystyle
		\nu\left( \frac{\rho_{gas}(t)}{\rho_{gas}(0)}\right) \qquad &\text{before GW}\\
		0\qquad &\text{after GW}\\
		\end{cases}
		\end{equation}
		where the star formation efficiency $\nu$, i.e. the proportionality coefficient between $\psi$ and the gas density, is assumed to be an increasing function of the galactic mass, in accordance to the prescription of the  ``inverse wind model'' \citep{matteucci1994,matteucci1998}.
	\item $\varphi(m)$ is the initial mass function (IMF); the different parameterizations used are detailed in the following sections.
\end{itemize}

\subsection{Energetics}\label{sec:energetics}
As previously mentioned, the code takes into account the effect of both Type I and  core-collapse (CC) SNe.
In particular, we assume a single degenerate scenario \citep{whelan1973} for Type Ia SNe (a C-O white dwarf accreting material from a red giant companion), so that the corresponding rate is given by \citep{greggio1983,matteucci1986,matteucci2001}:
\begin{equation}\label{SNI_rate}
	R_{SNIa} = A\,\int_{M_{B_m}}^{M_{B_M}}dM_B\,\varphi(M_B)\,\int_{\mu_m}^{0.5}\,f(\mu)\,\psi(t-\tau_m)\,d\mu
\end{equation}
where $M_B$ is the total mass of the binary system, assuming values - as previously mentioned - in the range $M_{B_m} - M_{B_M} (3-16 M_{\odot})$. The parameter $\mu\equiv M_2/M_B$ is the mass fraction of the secondary star (the originally least massive one) with respect to the total mass of the binary system, and we assume for it a distribution given by:
\begin{equation}
	f(\mu)=2^{\gamma+1}(\gamma + 1)\,\mu^{\gamma}
\end{equation}
where the best-fitting value of the free parameter $\gamma$ is found to be $\gamma=2$.\\
Finally, A is a free parameter, representing the fraction in the IMF of binary systems with the right properties to give rise to Type Ia SNe; its value is constrained in order to reproduce the present-day observed value \citep{cappellaro1999}.\\
As for CC SNe (II, Ib, Ic), their rate is given by:
\begin{equation}\label{SNII_rate}
	\begin{split}
	R_{cc} &= (1-A)\,\int_{8}^{16}dm\,\varphi(m)\,\psi(t-\tau_m) + \\
	& + \int_{16}^{M_{WR}}\,dm\,\varphi(m)\,\psi(t-\tau_m) + \\
	& + \int_{M_{WR}}^{M_{U}}\,dm\,\varphi(m)\,\psi(t-\tau_m) + \\
	& + \alpha_{Ib/c}\,\int_{12}^{20}\,\varphi(m)\,\psi(t-\tau_m)
	\end{split}
\end{equation}
where the first integral refers to the upper mass end of the third integral in the CEQ (single stars in the mass range $8-16\,M_{\odot}$), while the second one refers to the lower mass end in the last CEQ integral. $M_{WR}$ is the lower mass for a Wolf-Rayet star, i.e. the largest mass star ending its life as a  CC SN.\\
The last two integrals provide the Type Ib/c SNe rates, and account for single stars with masses larger than $M_{WR}$ (corresponing to the upper mass end in the last CEQ integral) and massive binary systems made of stars with masses in the range $12\leq m/M_{\odot}\leq20$, respectively.\\
$M_{U}$ is the mass limit assumed for the IMF, while $\alpha_{Ib/c}$ is a free parameter, representing the fraction of stars in the considered mass range which can actually produce Type Ib/c SNe.\\
Following \cite{cioffi1988}, only a few percent of the initial $\approx 10^{51}erg$ provided by core collapse SNe is actually injected into the ISM, due to cooling by metal ions; on the other hand, Type Ia SNe are assumed to contribute with their whole energy output, since they occur on a longer timescale in an ISM already heated by CC SNe \citep{recchi2001}. Overall, the efficiency of energy release averaged on both types is assumed to be $\approx 20\%$, following \cite{pipino2002}.\\
Once the SN energy input has been obtained, all is left to do is to determine the potential binding energy in the gas in each considered shell; this is done by integrating the equation
\begin{equation}
	E^{i}_{bin}(t) = \int_{R_i}^{R_{i+1}}\,dL(R)
\end{equation}
where $L(R)$ is the work required to bring a mass $dm=4\,\pi\,R^2\,\rho_{gas}\,dR$ from the shell radius $R_{i}$ to infinity \citep{martinelli1998}.\\
In order to calculate this integral, a model providing the baryonic and dark matter within a radius R is needed. For the baryonic (stars plus gas) matter, a \cite{jaffe1983J} profile is assumed:
\begin{equation}\label{eq_jaffe_profile}
	F_{bar}\propto\frac{r/r_0}{1+r/r_0}
\end{equation}
where $r_0 = R_{eff}/0.763$; as for dark matter (DM), it is assumed to be distributed in a diffuse halo, with a characteristic scale length ten times larger than the effective radius \citep[$R_{DM}=10\,R_{eff}$-][]{matteucci1992}, and with a profile taken from \cite{bertin1992}. The code computes the evolution with time of these quantities, and the galactic wind starts at the time $t_{GW}$ for which the following equality holds:
\begin{equation}
	E_{th}^{i}(t_{GW}) = E^{i}_{bin}(t_{GW})
\end{equation}

\subsection{Stellar yields}
Regarding the stellar yields, needed to correctly model nucleosynthesis, we adopted the same ones as used in \cite{P04}; specifically, for
\begin{itemize}
	\item single low and intermediate-mass stars (LIMS: $0.8<M/M_{\odot}<8$): metallicity-dependent yields by \cite{vandenhoek1997};
	\item SNe Ia in the single degenerate scenario, i.e. C-O white dwarfs in binary systems accreting material from a companion and exploding via C-deflagration upon reaching the Chandrasekar mass: yields by 
	\cite{Iwamoto1999};
	\item massive stars ($M>8\,M_{\odot}$): yields by \cite{thielemann1996} (hereafter, TNH96).
\end{itemize}
We have chosen this particular set of yields in order to compare our results with those of PM04.

\subsection{Comparison with data}

Unlike spirals, where we can directly estimate the chemical composition in the ISM \citep{pilyugin2001oxygen,kewley2002using,pettini2004iii,stasinska2004abundance,tremonti2004origin,pilyugin2005oxygen,kewley2008metallicity}, elliptical galaxy observations generally provide information about stellar abundances.\\
Specifically, the usual procedure consists in fitting the absorption features observed in the galactic integrated spectra to spectro-photometric models generated by evolutionary population synthesis (EPS) codes  \citep{bruzual1983,renzinibuzzoni1986,chiosi1988,buzzoni1989,charlotbruzual1991,bruzualcharlot1993,worthey1994old,tantalo1996,maraston1998,brocato2000,bruzual2003stellar,maraston2003stellar,thomas2003stellar,peletier2013stellar}, which allows us to constrain the age and composition of the galaxy; these spectral features, however, result from the contribution of many different stellar populations, so that the final abundance estimate is actually reflecting the abundance of the stellar population dominating the visual light.
\begin{figure*}
	\begin{subfigure}{.49\textwidth}
		\centering
		\includegraphics[width=1\linewidth]{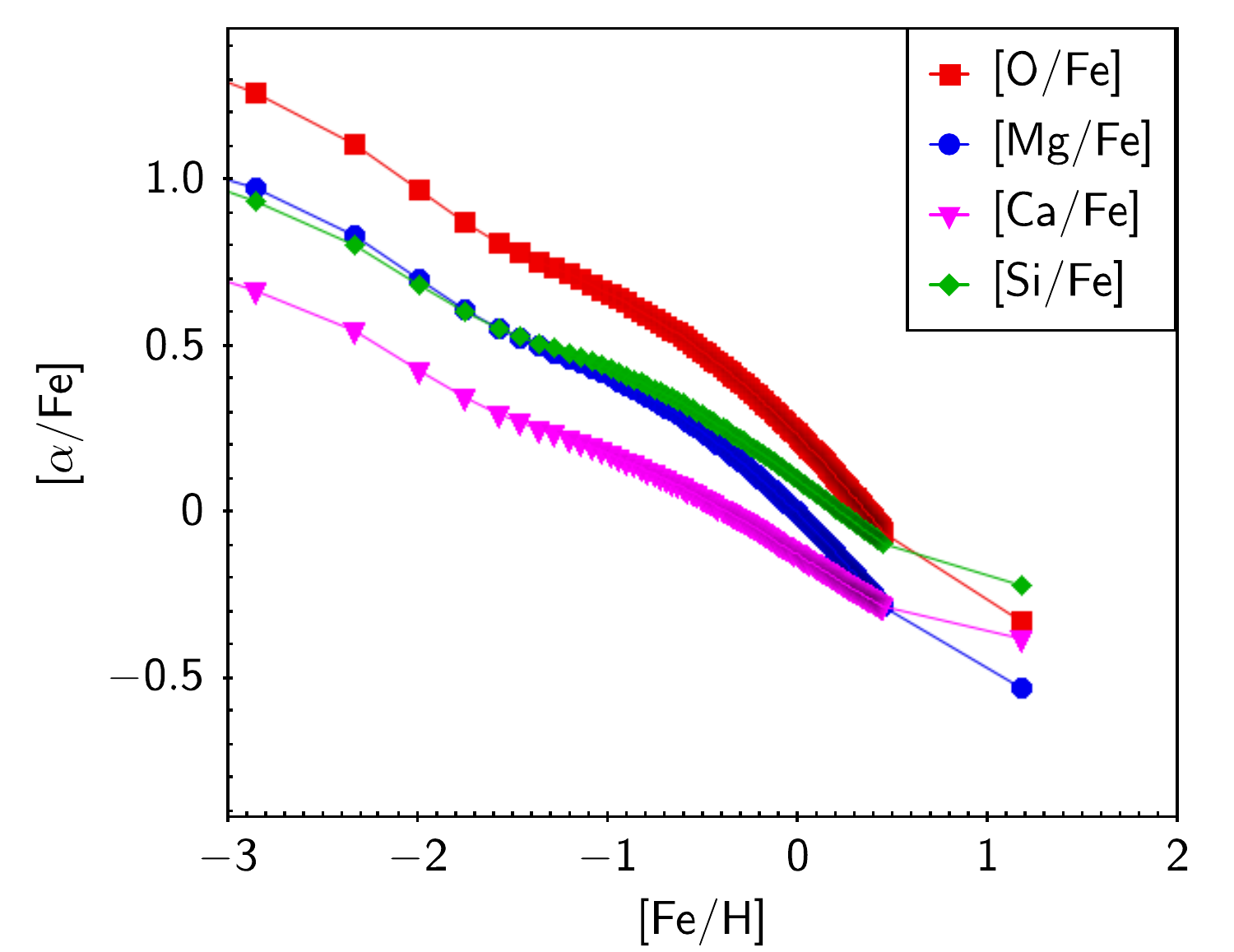}
		\end{subfigure}
	\begin{subfigure}{.49\textwidth}
		\centering
		\includegraphics[width=1\linewidth]{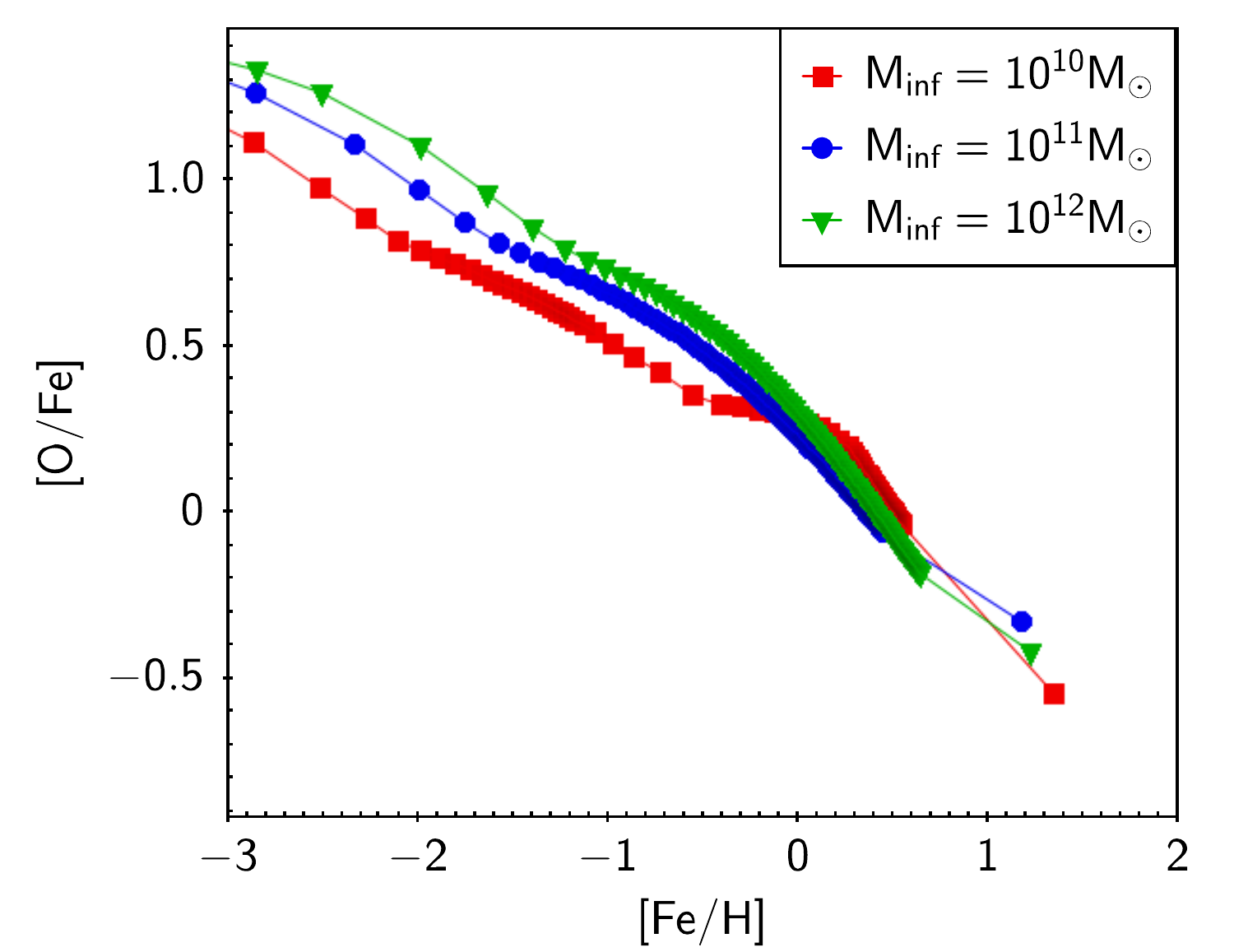}
	\end{subfigure}
\caption{Output of the chemical evolution code. Left: theoretical abundance ratios in the ISM of [O/Fe] (line with overlying squares), [Mg/Fe] (line with overlying circles), [Si/Fe] (line with overlying rhombus), [Ca/Fe] (line with overlying triangles) as functions of [Fe/H] in the core of a $10^{11}\,M_{\odot}$ galaxy. Right: theoretical [O/Fe] abundance ratio in the ISM as functions of [Fe/H] for the core of galaxies with $10^{10} M_{\odot}$ (squares), $10^{11} M_{\odot}$ (circles) and $10^{12} M_{\odot}$ (triangles) initial infall masses. The parameters used to obtain this output are the same as in Model 01a (see Table \ref{table:models01_parameters}).}\label{fig:model_output}
\end{figure*}
On the other hand, our chemical evolution code provides the chemical abundance in the ISM as a function of time for 21 different chemical elements; for reference, Fig. \ref{fig:model_output} shows the abundances of different $\alpha$-elements in the core ($0-0.1\,R_{eff}$) of a galaxy with a $10^{11}\,M_{Sun}$ stellar mass (left panel), and the [O/Fe] ratio in the cores of galaxies of different total masses (right panel).\\
In order to compare the outputs of our models with the observed averaged stellar abundances of the galaxies in the dataset, we first computed for each model the average stellar abundances at the present time. This can be done by averaging either on luminosity or mass \citep{matteucci1998}.\\
Since indices are usually obtained by weighting on the V-band luminosity \citep{arimoto1987, matteucci1998}, the more physically correct way to proceed is by using luminosity-weighted abundances; however, it has been shown that the results are not significantly different in massive galaxies \citep{matteucci1998}, so that in this work we always used only mass-weighed quantities, which were readily obtainable from our models.\\
The mass-weighted abundance of the element X is defined as \citep{pagel1975, matteucci2012}:
\begin{equation}
<X/H>_{mass} \equiv \frac{1}{M_0} \, \int_{0}^{M_0}Z(M)\,dM
\end{equation}
where $M_0$ is the total mass of stars ever born contributing to light at the present time; an alternative, equivalent formulation is given in \cite{pagel1997}
\begin{equation}
<X/H>_{mass}(t) \equiv \frac{ \int_{0}^{t}dt'\,(X/H)(t')\,\psi(t')  }{\int_{0}^{t'}\,dt'\,\psi(t')}
\end{equation}
where $\psi(t)$ is, as usual, the SFR.\\
Once the average has been determined, it can be converted to spectral indices to compare with observations by using calibration relations. In this work, we adopted the one derived from 
\cite{tantalo1998}, taking into account the $[Mg/Fe]$ enhancement:
\begin{subnumcases}{}
	Mg_2 & = 0.233 + 0.217\,[Mg/Fe] + \\ \nonumber & + (0.153+0.120[Mg/Fe])[Fe/H]\\
	<Fe> & = 3.078 + 0.341\,[Mg/Fe] + \\ \nonumber & + (1.654-0.307[Mg/Fe])[Fe/H]
\end{subnumcases}

\section{Results}\label{sec:results}

In this section, we describe the results obtained from the comparison of the output of our models with the objects in the catalog described in Section \ref{sec:catalog_T10}.\\
Specifically, we ran models for various values of the initial infalling gas mass, star formation efficiency, infall timescale and IMF.

\subsection{Star formation efficiency variation}\label{sec:basic_params_variations}

\begin{figure*}
	\includegraphics[width=1\linewidth]{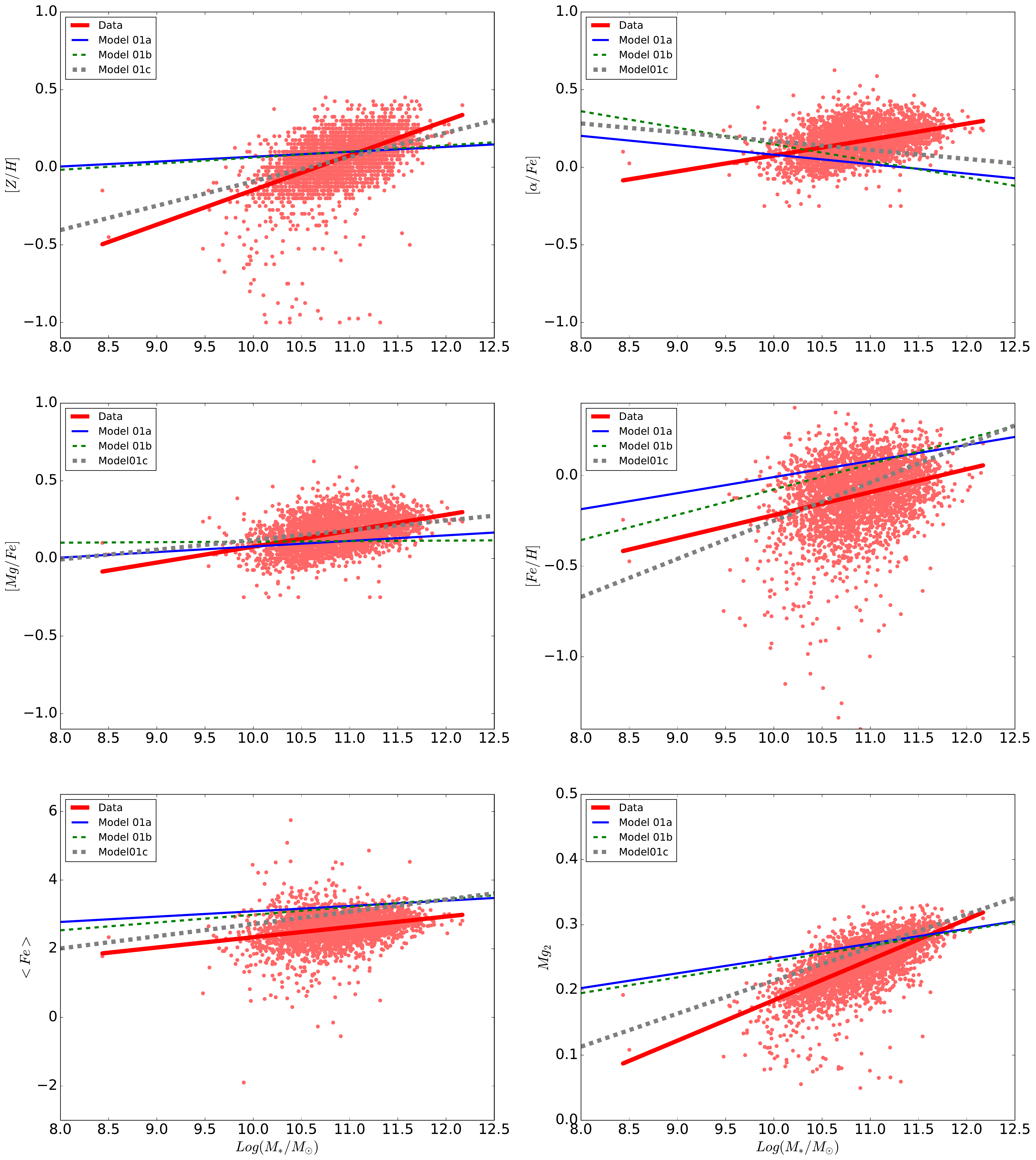}
	\caption{Comparison between the observed abundance patterns and the ones derived from Models 01. The models  are described in Section \ref{sec:basic_params_variations}, and their features summarized in Table \ref{table:models01_parameters}. The red dots represent galaxies in the catalog, while the lines indicate linear fit to data (thick solid line) and to Models 01a (thin solid line), 01b (thin dashed line) and 01c (thick dashed line) respectively.}\label{fig:model01}
\end{figure*}
In a first series of tests, we compared the observed data to different Models (i.e., sets of galaxies characterized by different variation of their physical parameters with the total mass), with the following assumptions:
\begin{itemize}
	\item initial infall masses in the $5\times10^9-10^{12}\,M_{Sun}$ range;
	\item effective radius increasing with the galactic mass;
	\item according to the inverse wind model prescription, increasing star formation efficiency $\nu$ and decreasing infall timescale $\tau$ for more massive galaxies.
\end{itemize}
In Table \ref{table:models01_parameters}, we report the values of the parameters used in creating the model galaxies; column 1 indicates the different sets of models, while columns 2 to 7 report the initial infall mass, the effective radius (the final radius achieved once the collapse is over), the star formation efficiency, the infall timescale, the time of the galactic wind onset calculated by the code and the assumed IMF for the different model galaxies, respectively. Notice that, in agreement with the downsizing formation scenario, the galactic wind appears earlier as the total mass increases.\\
As a starting point, in Model 01a we adopted the same parameters as in the best model of PM04. For galaxies with masses different from the ones analyzed in PM04, we obtained $R_{eff}$, $\nu$ and $\tau$ by interpolating between the known values.
\begin{table*}
	\centering
	\caption[Parameters for Models 01.]{Parameters used to create the three sets of models described in Section \ref{sec:basic_params_variations}. Within each set of models (column 1), we modified the star formation efficiency $\nu$, the infall timescale $\tau$ and the effective radius $R_{eff}$ (columns 4, 5 and 3 respectively) for different values of the initial infall mass $M_{inf}$ (column 2), and we always assumed a Salpeter IMF (column 7). Column 6 reports the time of the onset of the galactic wind in the model galaxies calculated by the code.}\label{table:models01_parameters}
	\begin{tabular}[c]{lcccccr}
		\hline
		Model                & $M_{inf}/M_{Sun}$ & $R_{eff}\,(kpc)$ & $\nu\,(Gyr^{-1})$ & $\tau\,(Gyr)$ & $t_{GW}\,(Gyr)$& IMF \\
		\hline
		\multirow{7}{*}{01a} & $5\times 10^{9}$  &  0.44  &   2.61  & 0.50  & 1.63 & Salp \\
		                     & $1\times 10^{10}$ &  1.00  &   3.00  & 0.50  & 1.49 & Salp \\
		                     & $5\times 10^{10}$ &  1.90  &   6.11  & 0.46  & 0.92 & Salp \\
		                     & $1\times 10^{11}$ &  3.00  &  10.00  & 0.40  & 0.56 & Salp \\
		                     & $5\times 10^{11}$ &  6.11  &  15.33  & 0.31  & 0.58 & Salp \\
		                     & $1\times 10^{12}$ & 10.00  &  22.00  & 0.20  & 0.45 & Salp \\
		\hline
		\multirow{7}{*}{01b} & $5\times 10^{9}$   &  0.44  &   0.89   & 0.50 & 3.72 &  Salp \\
	                         & $1\times 10^{10}$  &  1.00  &   1.00   & 0.50 & 3.34 &  Salp \\
		                     & $5\times 10^{10}$  &  1.90  &   1.89   & 0.46 & 2.17 &  Salp \\
		                     & $1\times 10^{11}$  &  3.00  &   3.00   & 0.40 & 1.52 &  Salp \\
		                     & $5\times 10^{11}$  &  6.11  &  11.44   & 0.31 & 0.67 &  Salp \\
		                     & $1\times 10^{12}$  & 10.00  &  22.00   & 0.20 & 0.45 &  Salp \\
		\hline
		\multirow{7}{*}{01c} & $1\times 10^{10}$  &  1.00  &   0.10   & 0.50  & 13.64 &  Salp \\
		                     & $5\times 10^{10}$  &  1.90  &   4.50   & 0.46  & 1.15  &  Salp \\
		                     & $1\times 10^{11}$  &  3.00  &  10.00   & 0.40  & 0.56  &  Salp \\
		                     & $5\times 10^{11}$  &  6.11  &  50.00   & 0.31  & 0.33  &  Salp \\
		                     & $1\times 10^{12}$  & 10.00  &  100.00  & 0.20  & 0.24  &  Salp \\
		\hline
	\end{tabular}
\end{table*}
In Figure \ref{fig:model01} we compare data (red points) and the predictions of our models. For each Model, we obtained the chemical properties for all the galaxies in the set, and performed a linear fit between these points to highlight the trends with galactic mass; this comparison shows how, in general, the chemical abundance patterns in Model 01a provide a poor agreement with the ones observed in the data. Both the total metallicity [Z/H] and $Mg_{2}$ index variations are considerably flatter than the observed ones, while the $[\alpha/Fe]$ ratio, defined by inverting equation \ref{eq:AFE_definition}, even presents an inverse trend with respect to data and to the expected one. On the other hand, it is worth to point out that when considering the $[Mg/Fe]$ ratio provided by the code as a proxy for $[\alpha/Fe]$ (third panel of Fig. \ref{fig:model01}), the trend of the latter increasing in more massive galaxies is preserved, even if the slope is generally still too shallow with respect to the data.
In the fourth panel of Fig. \ref{fig:model01}, we compared the $[Fe/H]$-mass relationship as predicted from our model with the one derived from the data by inverting equation \ref{eq:AFE_definition} (the $[Fe/H]$ ratio is not provided in the dataset). A small positive trend of the $[Fe/H]$ ratio with mass is apparent in the data, and its slope is reproduced by the model with reasonable agreement, although the $[Fe/H]$ values are slightly overestimated. Taken this into account, it appears that the flatness of the $[\alpha/Fe]$ trend with mass in the model can be attributed to the $\alpha$-elements abundance not increasing enough in massive galaxies.\\
Given the discrepancies between models and data, we tested a few other parameter configurations.\\
In Model 01b, in an attempt to increase the slope of the [Z/H] and $Mg_2$ curves, we decreased the star formation efficiency in low-mass galaxies; however, as shown in figure \ref{fig:model01}, this modification proved to be almost ineffective, producing little to no impact.\\
Finally, in Model 01c we further increased the difference in star formation efficiency between low and high mass galaxies, as again summarized in Table \ref{table:models01_parameters}. Though slightly improving the agreement with data for the total metallicity [Z/H] and the $Mg_2$ index in the Tantalo calibration, this variation generally provided even worse results for the $[\alpha/Fe]$ ratio.

\subsection{IMF variation}\label{sec:IMF_variations}
The results described in the previous section show how the attempts of reproducing the data trends by modifying the star formation efficiency and the infall timescale in the models proved to be unsuccessful; for this reason, we decided to test the effect of varying the IMF used in the models.\\
The topic of the IMF in elliptical galaxies has been long discussed in literature.
Analyzing the absorption line spectra of a sample of 38 early type galaxies and the bulge of M31 by their population synthesis models, \cite{conroyvandokkum2012} were able to obtain constraints on the IMF and the $(M/L)$ ratios of the individual objects, finding evidence for bottom heavier IMFs in galaxies with higher central velocity dispersions and $[Mg/Fe]$.\\
Results pointing to the same direction have been obtained from kinematics and gravitational lensing \citep{auger2010,grillo2010,treu2010,cappellari2012,spiniello2012} and from scaling relations and
global models of galaxies and dark matter \citep{dutton2011,dutton2012,dutton2013}.
\begin{figure*}
	\includegraphics[width=1\linewidth]{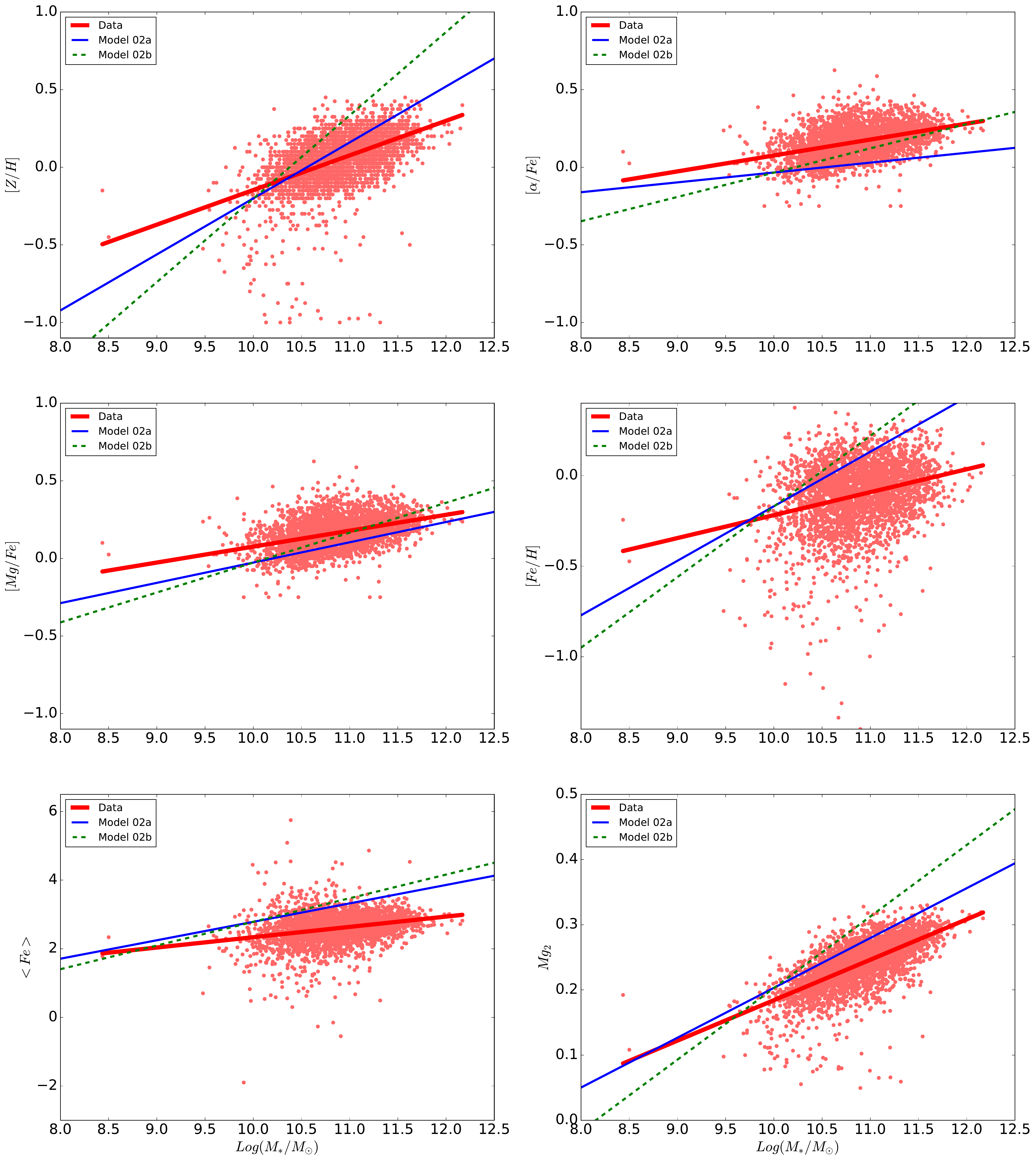}
	\caption{Comparison between the observed abundance patterns and the ones derived from Models 02. The models are described in Section \ref{sec:IMF_variations}, and their features summarized in Table \ref{table:models02_parameters}. The red dots represent galaxies in the catalog, while the lines indicate the linear fit to data (thick solid line) and to Models 02a (thin solid line) and 02b (thin dashed line) respectively.}\label{fig:model02}
\end{figure*}
However, in order to reproduce the increase of $[\alpha/Fe]$ ratios with stellar mass one has to assume an IMF becoming top heavier at higher masses \citep{matteucci1997}. Specifically, the IMF we investigated are:
\begin{itemize}
	\item \textbf{Scalo (1986) IMF}: we used the approximate expression adopted in \cite{chiappini1997}:
	\begin{equation}
	\varphi(m)\,\propto
	\begin{cases}
	m^{-2.35}\qquad & 0.1 \leq m/M_{\odot} < 6\\
	m^{-2.7}\qquad & 6 \leq m/M_{\odot} \leq 100\\
	\end{cases}
	\end{equation}
	\item \textbf{Salpeter (1955) IMF}, which is a simple power-law:
	\begin{equation}
	\varphi(m)\,\propto m^{-2.35}\qquad 0.1 \leq m/M_{\odot} < 100\\
	\end{equation}
	\item \textbf{Chabrier (2003) IMF}:
	\begin{equation}
	\varphi(m)\,\propto
	\begin{cases}
	e^{ - \frac{(Log(m) - Log(0.079))^2}{2(0.69)^2}}     \qquad & 0.1 \leq m/M_{\odot} < 1\\
	m^{-2.2}\qquad & 1 \leq m/M_{\odot} \leq 100\\
	\end{cases}
	\end{equation}
	\item \textbf{Arimoto \& Yoshii (1987) IMF}:
	\begin{equation}
	\varphi(m)\,\propto m^{-1.95}\qquad 0.1 \leq m/M_{\odot} < 100\\
	\end{equation}	
\end{itemize}
\begin{figure*}
	\centering
	\includegraphics[width=.7\linewidth]{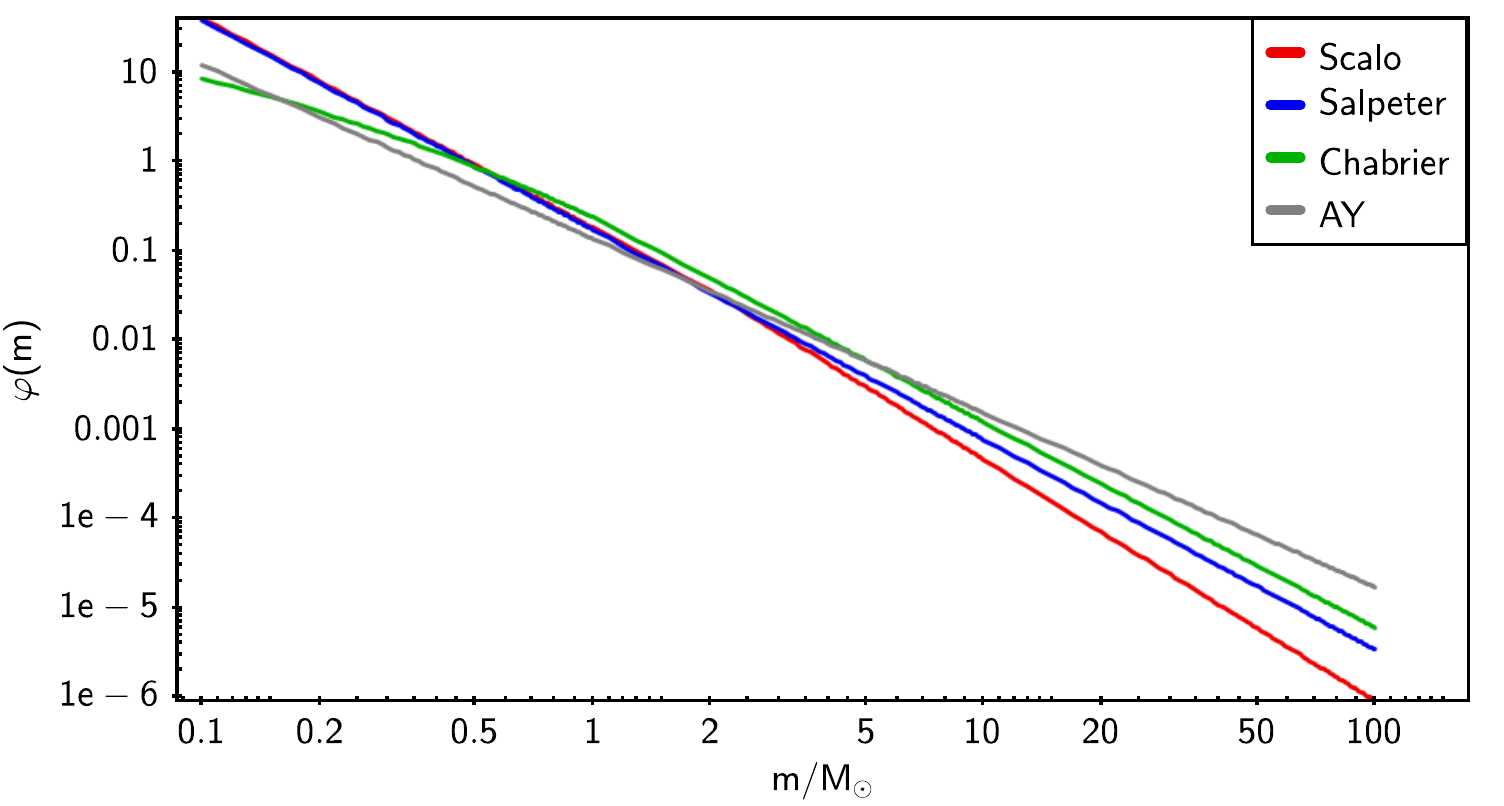}
	\caption{Comparison of the different IMFs used for the Models 02.}
	\label{fig:IMFs_comparison}
\end{figure*}
\begin{table*}
	\centering
	\caption[Parameters for Models 02.]{Parameters used to create the sets of models described in Section \ref{sec:IMF_variations}. The star formation efficiency $\nu$, the infall timescale $\tau$ and the effective radius $R_{eff}$ (columns 4, 5 and 3 respectively) are the same as used in Models 01a, but the assumed IMF (column 7) varies for different values of the initial infall mass $M_{inf}$ (column 2). Column 6 reports the time of the onset of the galactic wind in the model galaxies calculated by the code.}\label{table:models02_parameters}
	\begin{tabular}[c]{lcccccr}
		\hline
		Model                & $M_{inf}/M_{Sun}$ & $R_{eff}\,(kpc)$ & $\nu\,(Gyr^{-1})$ & $\tau\,(Gyr)$ & $t_{GW}\,(Gyr)$& IMF \\
		\hline
		\multirow{7}{*}{02a} & $5\times 10^{9}$  &  0.44  &   2.61  & 0.50  & 1.59 &  Scalo \\
	                         & $1\times 10^{10}$ &  1.00  &   3.00  & 0.50  & 1.37 &  Scalo \\
		                     & $5\times 10^{10}$ &  1.90  &   6.11  & 0.46  & 0.92 &  Salp \\
		                     & $1\times 10^{11}$ &  3.00  &  10.00  & 0.40  & 0.56 &  Salp \\
		                     & $5\times 10^{11}$ &  6.11  &  15.33  & 0.31  & 0.52 &  Chabrier \\
		                     & $1\times 10^{12}$ & 10.00  &  22.00  & 0.20  & 0.41 &  Chabrier \\
		\hline
		\multirow{7}{*}{02b} & $5\times 10^{9}$  &  0.44  &   2.61  & 0.50  & 1.59 &  Salp \\
		                     & $1\times 10^{10}$ &  1.00  &   3.00  & 0.50  & 1.37 &  Salp \\
		                     & $5\times 10^{10}$ &  1.90  &   6.11  & 0.46  & 0.92 &  Salp \\
		                     & $1\times 10^{11}$ &  3.00  &  10.00  & 0.40  & 0.56 &  Salp \\
		                     & $5\times 10^{11}$ &  6.11  &  15.33  & 0.31  & 0.51 &  AY \\
		                     & $1\times 10^{12}$ & 10.00  &  22.00  & 0.20  & 0.41 &  AY \\
		\hline
	\end{tabular}
\end{table*}
Varying the IMF from Scalo 1986 to \cite{chabrier2003} has greatly improved the agreement of model predictions with data.\\
In particular, Model 02a (see Table \ref{table:models02_parameters}) has provided the best results, being able to simultaneously reproduce the observed trends for all the considered quantities. On the other hand, Model 02b, which contains the same IMF variation adopted in \cite{matteucci1994}, has provided overestimated values for all the chemical properties in high mass galaxies, which can be attributed  to the excessive numbers of massive stars produced by assuming the \cite{arimoto1987} IMF.\\
A comparison with Figure \ref{fig:model01} shows how, in this case, the slope of the $[Fe/H]$-mass relationship predicted by our models is steeper than the one derived from equation \ref{eq:AFE_definition} for the data; in spite of this, the top-heavier IMF, and therefore the larger number of produced massive stars, leads to an increased production of $\alpha$-elements, thus yielding a positive trend of the $[\alpha/Fe]$ ratio with mass.

\subsection{IGIMF}\label{sec:IGIMF_variations}

As a last step, we decided to test the effect of the integrated galactic initial mass function \citep[IGIMF - see ][]{kroupa2003,weidner2005,recchi2009,recchi2014,weidner2010,vincenzo2015} on the  predicted abundance patterns.\\
The idea at the basis of the IGIMF theory is that star formation is expected to take place mostly in embedded clusters \citep{ladalada2003}. Within each cluster, masses of new-born stars are indeed distributed following a canonical IMF $\varphi(m)\propto m^{-\alpha}$, while young embedded star clusters themselves are assumed to follow a mass function (embedded cluster mass function; ECMF).\\
\begin{figure*}	
	\centering
	\includegraphics[width=.7\linewidth]{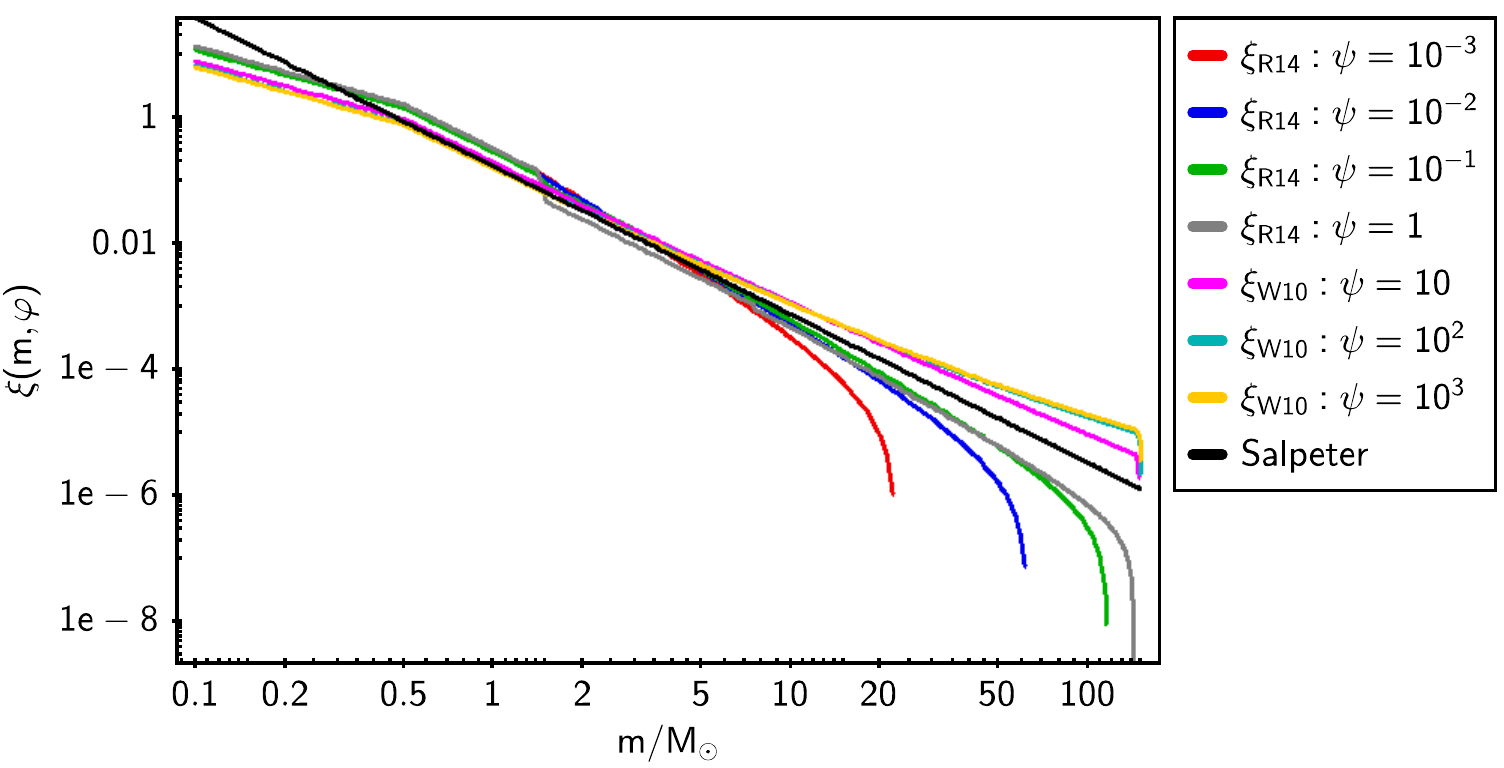}
	\caption{Comparison between the IGIMF for different SFRs and a canonical Salpeter IMF.}
	\label{fig:IGIMF_comparisons}
\end{figure*}
The ECMF has been found to be \citep{zhang1999,hunter2003,ladalada2003,recchi2009} a simple power-law
\begin{equation}
	\xi_{ecl} \propto M_{ecl}^{\beta}
\end{equation} 
where the index $\beta$ generally assumes values $\approx 2$.\\
This influences the resulting stellar IMF due to two concurring main factors:
\begin{figure*}
	\includegraphics[width=1\linewidth]{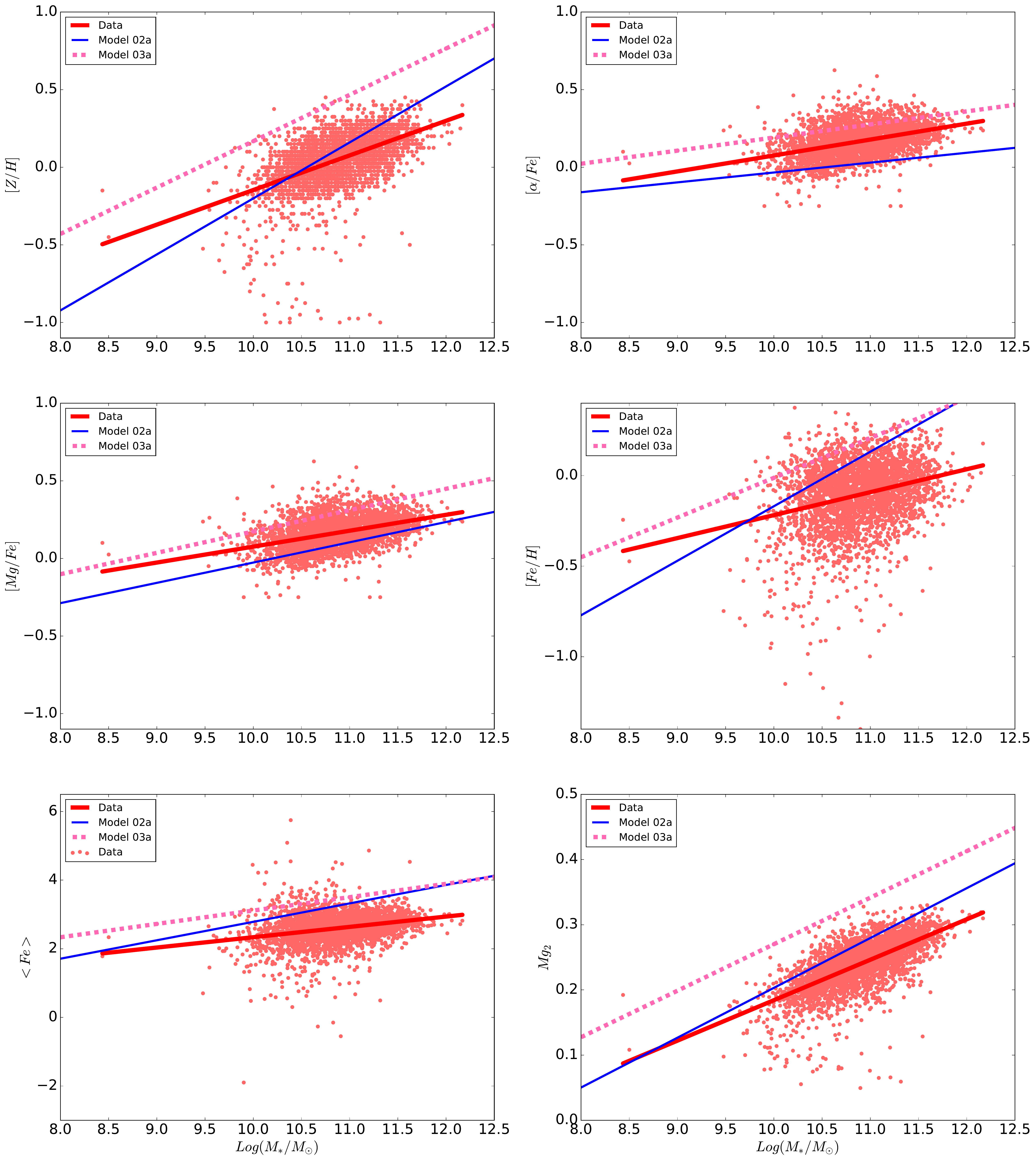}
	\caption{Comparison between the observed abundance patterns and the ones derived from Models 02a and Model 03a, which assumes an IGIMF. The red dots represent galaxies in the catalog, while the lines indicate the linear fit to data (thick solid line) and to Models 02a (thin solid line) and 03a (thick dashed line) respectively.}\label{fig:models03}
\end{figure*}
\begin{enumerate}
	\item the upper mass limit of the ECMF, i.e. the most massive star forming embedded cluster, has been found to depend on the SFR of the galaxy, in the sense that it increases in more star-forming environments \citep{weidner2004,bastian2008,weidner2011};
	\item the maximum stellar mass which can be formed within each embedded star cluster increases with the mass of the cluster itself \citep{weidner2010,weidner2011}.
\end{enumerate}
Therefore, more massive clusters are formed in galaxies with higher SFRs, and within these clusters stars with larger masses are produced.\\
\begin{table*}
	\centering
	\caption[Parameters for Models 03.]{Parameters used to create the sets of models described in Section \ref{sec:IGIMF_variations}. The star formation efficiency $\nu$, the infall timescale $\tau$ and the effective radius $R_{eff}$ (columns 4, 5 and 3 respectively) are the same as used in Models 01a and 02a, but we are assuming an IGIMF. Column 6 reports the time of the onset of the galactic wind in the model galaxies calculated by the model.}\label{table:models03_parameters}
	\begin{tabular}[c]{lcccccr}
		\hline
		Model                & $M_{inf}/M_{Sun}$ & $R_{eff}\,(kpc)$ & $\nu\,(Gyr^{-1})$ & $\tau\,(Gyr)$ & $t_{GW}\,(Gyr) $ & IMF \\
		\hline
		\multirow{7}{*}{03a} 	& $5\times 10^{9}$  &  0.44  &   2.61  & 0.50  & 1.60 &  IGIMF \\
								& $1\times 10^{10}$ &  1.00  &   3.00  & 0.50  & 1.43 &  IGIMF \\
								& $5\times 10^{10}$ &  1.90  &   6.11  & 0.46  & 0.96 &  IGIMF \\
								& $1\times 10^{11}$ &  3.00  &  10.00  & 0.40  & 0.53 &  IGIMF \\
								& $5\times 10^{11}$ &  6.11  &  15.33  & 0.31  & 0.55 &  IGIMF \\
								& $1\times 10^{12}$ & 10.00  &  22.00  & 0.20  & 0.44 &  IGIMF \\
		\hline
	\end{tabular}
\end{table*}
The resulting IGIMF is then obtained by considering all the IMFs in all of the embedded clusters, i.e. it is defined as \citep{weidner2011,vincenzo2015}:
\begin{equation}\label{eq:IGIMF_def}
	\begin{split}
	&\xi_{IGIMF}(m,t) \equiv\\ &\equiv\int_{M_{ecl}^{min}}^{M_{ecl}^{max}(\psi(t))}\,\varphi(m<m_{max}(M_{ecl}))\,\xi_{ecl}(M_{ecl})\,dM_{ecl}
	\end{split}
\end{equation}
and it is normalized in mass, so that \citep{vincenzo2015}:
\begin{equation}\label{eq:IGIMF_normalization}
	\int_{m_{min}}^{m_{max}}\,dm\,m\,\xi_{IGIMF}(m) = 1
\end{equation}
The IGIMF theory has been applied, with different prescriptions, both to systems with low star formation rates, like dwarf galaxies \citep{vincenzo2015}, and to starbursts \citep{weidner2010}; to account for the variation of the SFR, going from low values in the first time steps up to higher values throughout the evolution of the galaxies in our models, we decided to combine the two prescriptions.\\
In particular:
\begin{itemize}
	\item in the low star formation regime $(\lesssim 10\,M_{Sun}/yr)$, we have followed the approach by \cite{recchi2009}; in this case, the maximum mass limit of the embedded clusters is given by:
	\begin{equation}\label{eq:IGIMF_R14_Meclmax}
		Log(M_{ecl}^{max}) = 4.83 + 0.75\,Log\left( \psi(t)\right) 
	\end{equation}
	and the stellar IMF within each cluster is a two slope power law:
	\begin{equation}
	\varphi(m)=
		\begin{cases}
		A\,m^{-\alpha_1} \qquad &0.08\leq m/M_{Sun} \leq 0.5\\
		B\,m^{-\alpha_2} \qquad &0.5\leq m/M_{Sun} \leq m_{max}\\	
		\end{cases}
	\end{equation}
	where $\alpha_1=1.30$ and $\alpha_2=2.35$ (in the formulation by \cite{recchi2014}, $\alpha_2$ depends on the [Fe/H] abundance as well, but this is neglected in \cite{weidner2005} )
\item when the systems reached higher SFR values, we referred to the \cite{weidner2010} formulation. In this case, the dependence of the maximum cluster mass on the SFR is given by:
\begin{equation}\label{eq:IGIMF_W10_Meclmax}
	M_{ecl}^{max} = 8.5\times 10^4 \left( \psi(t)\right)^{0.75}  
\end{equation}
while the assumed stellar IMF is
\begin{equation}
	\varphi(m)= k
	\begin{cases}
		\left( \frac{m}{m_H}\right)^{-\alpha_0}\\
		\left( \frac{m}{m_H}\right)^{-\alpha_1}\\
		\left( \frac{m_0}{m_H}\right)^{-\alpha_1} \left( \frac{m}{m_1}\right)^{-\alpha_2}\\
		\left( \frac{m_0}{m_H}\right)^{-\alpha_1} \left( \frac{m_1}{m_0}\right)^{-\alpha_2}\left(\frac{m}{m_1}\right)^{-\alpha_3}\\
	\end{cases}
\end{equation}
with
\begin{equation*}
	\begin{split}
	\alpha_0 &= 0.30 \qquad 0.01 \leq m/M_{\odot} \leq 0.08 \\
	\alpha_1 &= 1.30 \qquad 0.08 \leq m/M_{\odot} \leq 0.50 \\
	\alpha_2 &= 2.35 \qquad 0.50 \leq m/M_{\odot} \leq 1.00 \\
	\end{split}
\end{equation*}
and, for $1.00 \leq m/M_{\odot} \leq m_{max}\\$
\begin{equation*}
	\alpha_3 =
	\begin{cases}
	2.35 \qquad & M_{ecl}<2\times10^5\,M_{\odot}\\
	-1.67 \times Log\left( \frac{M_{ecl}}{10^6\,M_{\odot}}\right) \qquad & M_{ecl}>2\times10^5\,M_{\odot}\\
	1 \qquad & M_{ecl}>1\times10^6\,M_{\odot}
	\end{cases}
\end{equation*}
\end{itemize}
Figure \ref{fig:IGIMF_comparisons} shows the IGIMF resulting from the combination of the two prescriptions for different values of the SFR, compared to a canonical Salpeter IMF.\\
The parameters used are summarized in Table \ref{table:models03_parameters}, and Figure \ref{fig:models03} shows the comparison between the results obtained with the IGIMF and the ones provided by Model 02a, which so far proved to be the best-fitting one.\\
Consistently with what we observed with the increasingly top-heavy IMF we presented in the previous section, the inclusion of the IGIMF had the effect of steepening the trends with mass produced by our models; as a matter of fact, the Models 03 were the ones with the best fitting slopes. On the other hand, the chemical abundances predicted by all of these models were always higher, to some extent, than what observed in the data, which is particularly noticeable when comparing the total metallicity $[Z/H]$ and the $Mg_2$ spectral index (as for the latter, it is worth stressing that, as for all spectral indices measurements, its value is greatly influenced by the assumed calibration relation).\\
Table \ref{table:linear_fits_coefficients} summarizes the linear fit coefficients obtained for data and all the models presented in the paper.

\section{Summary and conclusions}
We adopted a multi-zone model of chemical evolution for elliptical galaxies, taking into account SN feedback (including Type Ia, Ib, Ic and II SNe), and an initial fast infall episode leading to the formation of galaxies.\\
\begin{table*}
	\centering
	\caption{Coefficients of the linear fits $(y=mx+q)$ for data and all the models.}\label{table:linear_fits_coefficients}
	\begin{tabular}[c]{lcccccccccr}
		\hline
		& \multicolumn{2}{|c|}{[Z/H]} & \multicolumn{2}{|c|}{$[\alpha/Fe]$} & \multicolumn{2}{|c|}{$[Mg/Fe]$} & \multicolumn{2}{|c|}{$<Fe>$}  & \multicolumn{2}{|c|}{$Mg_2$} \\
		&  m   &    q &  m   &    q & m   &    q &  m   &    q & m   &    q \\
		\hline
		Data     & 0,223 & -2,377 &  0,103 & -0,95  & 0,103 & -0,950 & 0,301 & -0,671 & 0,062 & -0,436 \\
		Model01a & 0,032 & -0,248 &  0,061 & 0,688  & 0,036 & -0,282 & 0,155 &  1,537 & 0,023 &  0,020 \\
		Model01b & 0,039 & -0,332 & -0,107 & 1,215  & 0,003 &  0,074 & 0,227 &  0,722 & 0,024 &  0,001 \\
		Model01c & 0,157 & -1,660 & -0,057 & 0,739  & 0,063 & -0,509 & 0,357 & -0,849 & 0,050 & -0,293 \\
		Model02a & 0,361 & -3,808 &  0,064 & -0,669 & 0,131 & -1,333 & 0,537 & -2,588 & 0,076 & -0,561 \\
		Model02b & 0,537 & -5,573 &  0,157 & -1,600 & 0,193 & -1,954 & 0,689 & -4,110 & 0,110 & -0,895 \\
		Model03a & 0,299 & -2,819 &  0,084 & -0,652 & 0,138 & -1,202 & 0,389 & -0,781 & 0,071 & -0,443 \\
		\hline
	\end{tabular}
\end{table*}
We tested the predictions of the models against the dataset by \cite{thomas2010}, containing information on the chemical abundance patterns for $\approx 3000$ galaxies, which have been visually inspected and classified as ellipticals from a starting sample of SDSS DR4 galaxies. From the complete sample, we only considered the objects which did not show signs of recent star formation activity (the ``red sequence'' subset).\\
Besides the total metallicity $[Z/H]$ and the $[\alpha/Fe]$ ratio, we also derived an estimate of the <Fe> and $Mg_2$ Lick indices for our model galaxies by applying the \cite{tantalo1998} calibration to the predicted chemical abundances, and compared them to the corresponding values provided for the galaxies in the dataset.
In a first series of tests, we tried to reproduce the observed trend, namely the mass-metallicity and the $[\alpha/Fe]$ vs. mass relations, by only exploiting the downsizing formation hypothesis, and assuming a fixed IMF \citep{salpeter1955}. In the downsizing scenario, the  more massive galaxies have higher star formation efficiencies and smaller infall timescales; these hypotheses can in principle simultaneously account for the mass-metallicity and $[\alpha/Fe]$-mass relationships, as shown in \cite{P04}. The downsizing in star formation acts mainly on the time of the occurrence of a galactic wind with consequent quenching of star formation, which occurs first in more massive galaxies, if the efficiency of star formation is an increasing function of the galactic mass. This was first demonstrated by \cite{matteucci1994}, and indicated that the efficiency of star formation is a crucial parameter in galaxy evolution. In the case of the ellipticals, increasing enough the star formation efficiency leads to a faster development of the galactic wind in spite of the deeper potential well of the gas in the galaxy. This high efficiency can preserve the mass-metallicity relation and, at the same time, predicts that the bulk of stars in massive ellipticals show higher $[\alpha/Fe]$ ratios than smaller ones. The reason is that the wind occurs before in massive than in small objects, and this prevents SNe Ia, which are the major Fe producers, to pollute the gas with Fe. They will continue to produce Fe even after star formation has stopped, and this Fe will eventually end up in the intracluster medium.\\
\cite{P04} found a good agreement with the data available at that time, but the comparison with the new data has shown how the resulting agreement is not satisfying. Specifically, the chemical abundance trends with stellar mass provided by the models with a constant Salpeter IMF and downsizing in star formation, are generally flatter than the ones derived from the data.\\
In this sense, the agreement between the $[Fe/H]$-mass relationship predicted from the models and the one derived from the data showed how the flatness of the $[\alpha/Fe]$-mass relationship in the models could be mainly attributed to a insufficient production of $\alpha$-elements when only assuming the downsizing scenario.\\
We tried different combinations of the available parameters (e.g. star formation efficiency and infall timescale), stretching their variation with mass in an attempt to increase the slope of the chemical patterns observed in the models, but this exercise turned out to be mostly ineffective; the fit with data remains poor. For example, Model 01c, characterized by an extreme star formation efficiency variation, actually provides good agreement with most of the observed trends, but it fails in reproducing the $[\alpha/Fe]$ ratio as defined in equation \ref{eq:AFE_definition}, to the point of even predicting an inverse trend with mass (see Fig. \ref{fig:model01}). Moreover, as summarized in Table \ref{table:models01_parameters}, in this Model we had to stretch the $\nu$ variation to the point of assuming, for low mass galaxies, values even smaller than the ones usually assumed for dwarf galaxies \citep{vincenzo2014} and this is not a reasonable assumption. It should be noted  that part of the reason of the poor fit to the $[\alpha/Fe]$ vs. mass relation, is due to the adopted definition of the $\alpha$-elements as in equation \ref{eq:AFE_definition}, which is obtained by subtracting the Fe abundance from the global metallicity Z. This quantity is not correctly following the real behavior of $\alpha$-elements, since in the global metal content, Z, there are other abundant metals such as C and N, for example, which do not behave as $\alpha$-elements. In fact, if we plot just the [Mg/Fe] ratios versus mass the agreement with the data is reasonable.\\
Because of the poor agreement in the case of the fixed IMF, we tested models with different variable IMFs.\\
Contrary to the claims for a the need of a bottom heavy IMF in more massive galaxies, we followed the general idea that, in order to increase the $[\alpha/Fe]$ with galactic mass, more massive stars are needed in larger galaxies. For this reason, we computed models with IMFs becoming increasingly top-heavier in more massive galaxies, going from a Scalo (1986) for the less massive ellipticals, to a Salpeter \citep{salpeter1955} for the intermediate mass ones and then to a Chabrier \citep{chabrier2003} in more massive ones.\\ The combination of the IMF variation together with the downsizing formation scenario allowed us to successfully increase the range of values assumed by the chemical abundances in the models. We found the effect of the IMF (the transition from bottom to top heavier IMFs with the increasing total mass) is the dominant one, with the variation of the other parameters providing a smoothing effect between the different mass ranges. In other words, the IMF is the most effective parameter influencing the chemical evolution of galaxies.
Our assumed IMF variation (e.g. Scalo-Salpeter-Chabrier) provides the best simultaneous agreement between models and data for all the considered chemical properties ($[Z/H]$, $[\alpha/Fe]$, $[Mg/Fe]$, $<Fe>$ and $Mg_2$); although some fits to the data are not perfect (note in particular how these Models generally underestimate the $[\alpha/Fe]$ and $[Mg/Fe]$ ratios), the Model with the variable IMF, as described above, is the only one producing abundance patterns in reasonable good agreement with data. Another test, where we adopted an Arimoto-Yoshii \citep{arimoto1987} IMF in more massive galaxies (e.g. an IMF even flatter than the Chabrier one), showed that this IMF generally overestimate the metals in these galaxies.\\
Finally, we tested the inclusion in the models of the Integrated Galactic IMF (IGIMF), defined by integrating the canonic stellar IMF over the mass function of embedded clusters, within which star formation is assumed to take place. In this scenario, more massive clusters are formed in galaxies with higher SFRs, and within these clusters stars with larger masses are produced; as a consequence, the resulting IGIMF basically provides an IMF naturally becoming top heavier as the galactic mass increases. In order to determine the IGIMF values throughout the evolution of the galaxy, and the corresponding increase of the SFR values, we combined the two prescriptions of \cite{recchi2009} (for the initial evolutionary steps, characterized by a low star formation regime) and \cite{weidner2010} (for the later phases).\\
Generally, the slopes of the chemical relations produced by these Models show the best agreement with data relative to  all the explored solutions. However, a definite offset is always present, with the models overpredicting the various chemical abundances.\\
In conclusion, we like to stress that the comparison between abundances derived from indices in elliptical galaxies and models is an uncertain exercise, since the metallicity indices need to be converted into real abundances by means of suitable calibrations. The same occurs from a theoretical point of view, when one should transform the real abundances predicted by the models into indices. Even in this case, one should adopt a suitable calibration, but unfortunately different calibration can produce different results, as shown in \cite{P04}. Therefore, any conclusion based on such a comparison should be taken with care. So, we do not expect to reproduce the exact values of the abundances and of the abundance ratios in these galaxies, but rather the trends shown by such abundances. What we have shown here is that, in order to reproduce the observed slopes of the chemical relations, namely the mass-metallicy and the mass-$[\alpha/Fe]$ relation at the same time, one should assume a downsizing in star formation, implying that the most massive galaxies are the oldest, as well as a variation in the IMF favoring massive stars in more massive galaxies. In other words, our results show that a bottom-heavy IMF, favoring low mass stars relative to massive ones, is not likely to reproduce the observed chemical trends in ellipticals. In the future, further investigation should be carried out to eliminate the possibility of calibration biases, so to allow us to better compare model results with observations.\\

\section*{Acknowledgements}
CDM wishes to thank E. Spitoni for the precious help and the numerous suggestions. CDM and FM acknowledge research funds from the University of Trieste (FRA2016). FV acknowledges funding from the United Kingdom Science and Technology Facilities Council through grant ST/M000958/1. We also thank the anonymous referee for his useful comments, which improved the content and clarity of the paper.



\bibliographystyle{mnras}
\bibliography{bibliography.bib}




\appendix


\bsp	
\label{lastpage}
\end{document}